# Wasp - Waisted loop and Spin frustration in $Dy_{2-x}Eu_xTi_2O_7$ Pyrochlore

Prajyoti Singh[1], Arkadeb Pal[1], Vinod K. Gangwar[1], Prince K Gupta[1], Mohd. Alam[1], Surajit Ghosh[1], R. K. Singh[2], A. K. Ghosh[2] and Sandip Chatterjee[1,*]

[1]Department of Physics, Indian Institute of Technology (BHU), Varanasi – 221005, India

[2]Department of Physics, Banaras Hindu University, Varanasi – 221005, India

*Corresponding author's email address: schatterji.app@iitbhu.ac.in

**Abstract**

The Raman spectroscopy and AC and DC magnetization of $Dy_{2-x}Eu_xTi_2O_7$ have been investigated. In Raman Spectroscopy, the systematic shift in all phonon modes with Eu content in $Dy_{2-x}Eu_xTi_2O_7$ confirms that $Dy^{3+}$ ion is substituted by $Eu^{3+}$ ions. High concentration of Eu induces the dipolar exchange interactions and crystal-field interactions in $Dy_{2-x}Eu_xTi_2O_7$. Rich Eu content samples (x=1.8 and 1.9) show the existence of wasp-waisted hysteresis loop and that can be attributed to the coexistence of dipolar field and anisotropy exchange interaction. AC susceptibility shows two single ion spin freezing transitions corresponding to $Dy^{3+}$ and $Eu^{3+}$ ions respectively in x = 1.5, 1.8, 1.9 samples.

## 1. Introduction

Systems of interacting degrees of freedom possess competing magnetic interactions or geometrical frustration. Triangular magnetically interacting spins are incapable to order into a single magnetic ground state due to competing correlations between different lattice hosting these spins. These competing interactions are driven into macroscopically degenerate ground states at low temperature [1-3]. These exotic magnetic ground states are classified as classical dipolar spin ice [4, 5], quantum spin ice [6], spin slush [7], quantum spin liquid states [8, 9], spin glass [10-12] and order by disorder states [13-15]. These novel and riveting ground states are compiled by balancing the dipolar interactions, exchange interactions and strong crystal field effects. Geometrical frustrated pyrochlore structure (space group Fd-3m) has an empirical formula $A_2^{3+}B_2^{4+}O_7$ or $A_2B_2O_6O^{'}$ in which A-site (16d Wyckoff position) contains rare earth ions and B-site (16c) accommodates transition metal ions. Rare earth cation $A^{3+}$ is surrounded by eight oxygen anions [six O (48f) ions and two O′ (8b) ions] where as transition metal cation $B^{4+}$ is coordinated with six oxygen anion O (48f). Hence, ordered pyrochlore structure is defined as an arrangement of two interpenetrating networks of $BO_6$ octahedral and $A_2O′$ chains and thus, constituting a three-dimensional array of corner sharing tetrahedra [16]. Among all these frustrated systems, the spin ice state has its own forte. Study of materials with antiferromagnetic nearest neighbour interaction is typical for geometrical magnetic frustration. But for particular spin ice compounds [$Dy_2Ti_2O_7$ (DTO), $Ho_2Ti_2O_7$ (HTO), $Ho_2Sn_2O_7$ (HSO), $Dy_2Sn_2O_7$ (DSO) and $Dy_2Ge_2O_7$ (DGO)] in which ferromagnetic and dipolar interactions are frustrated, they also show strong frustration effect [4, 5, 17-27]. In spin ice materials, ferromagnetic interactions, dipole exchange interactions, strong single-ion Ising anisotropy and crystal field effects contrive the spins of rare earth ion to align directly towards (two spins) and directly away ( two spins) from the centre of the tetrahedra. This "two-in/two-out" arrangement of spins is similar to arrangement of hydrogen ions in water ice [17]. Although a long range ordered state has been suggested by theoretical calculation [28] to possess non zero-point entropy for these spin ice materials, experimentally, they show no long range ordering in zero magnetic field. The spins freeze into a non-equilibrium low temperature state around same zero point entropy as water ice [4, 29]. Recently, it has been observed that the dipole-dipole interactions in spin ice materials elevate to magnetic monopole [30-32].

$Dy_2Ti_2O_7$ is an acclaimed spin ice pyrochlore system in which $Dy^{3+}$ ions (having 7f $e^-$ with J = 15/2) reside on a lattice of corner sharing tetrahedra. The uniaxial anisotropy of these $Dy^{3+}$ ions which points along <111> axis begets the Ising type ground state doublet with a large gap of ~ 200 K below the first excited state [5, 21, 33]. Positive Curie-Weiss temperature ($\theta_{CW} \approx 1$ K) of DTO pyrochlorereported earlier, confirms the ferromagnetic interaction between the nearest neighbours [4]. It has been reported from the specific heat measurement of DTO compound that spin freezes out only below the $T_i$ = 4 K [24]. However, AC susceptibility study shows two novel dynamic spin freezing at $T_i$ ~ 4 K and $T_f$ ~ 16 K. The low temperaturetransitionat $T_i$ ~ 4 K is associated to spin freezing of ice rules [34, 35] and the high temperature spin dynamics ( at $T_f$ ~ 16 K) has been reported to originate from spin flip process [36, 37]. This spin relaxation mechanism of DTO shows a surprising classical to quantum double cross over process upon cooling as given flow chart below;

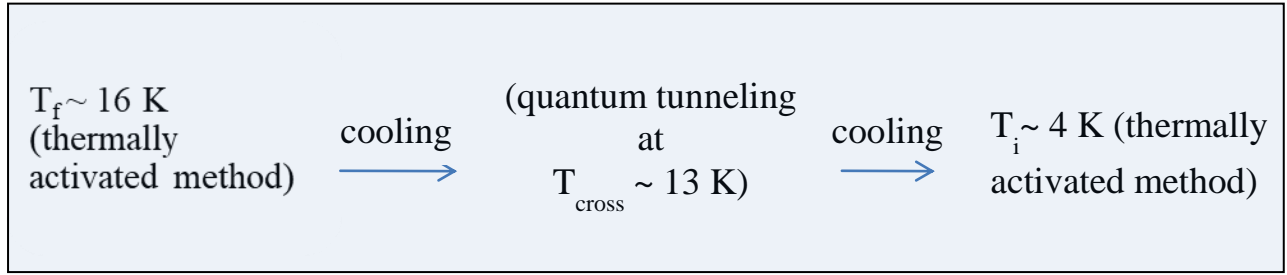

The development of spin correlations is thought to be the reason behind the cross over [34, 38]. Eventually, the presence of high temperature spin freezing ($T_f \sim 16$ K) makes DTO system more fascinating among all spin ice systems because all other spin ice compounds do not show such type of high temperature spin freezing. $Dy_2Ti_2O_7$ is renowned not only for magnetic equity but also for its electric conductivity [39], dielectric and magneto dielectric responses [40, 41].

Another frustrated pyrochlore system $Eu_2Ti_2O_7$ (ETO) exhibits antiferromagnetic interaction as it has negative Curie-Weiss temperature ($\sim \theta_{CW} = -1.35$ K) [42]. Although, ETO has a non magnetic ground state, it shows substantial magnetic moment and strong crystal field effect which develops single ion anisotropy parallel to <111> axis [43]. An exotic spin freezing transition around $T_s = 35$ K was found in AC susceptibility measurement of ETO compound, which is highest reported spin freezing transition so far in geometrically frustrated pryochlore systems. This spin freezing was assigned to single spin relaxation type process by dilution of $Y^{3+}$ ion in ETO [42].

Many reports are available on both (A and B) sides doping on $Dy_2Ti_2O_7$ to change the spin-configuration at low temperature. Substitution of $Dy^{3+}$ ions by non magnetic ($Y^{3+}$ or $Lu^{3+}$) ionsreveals the reemergence of high temperature spin freezing transition which shifts towards higher temperature side with increasing doping concentration [34]. Not only non magnetic diluents [34] but also magnetic elements (like $Tb^{3+}$ ion) show drastic change in the ground state of $Dy_xTb_{2-x}Ti_2O_7$ pyrochlore [44]. Some reports on Ti site doped compounds of DTO are also available [45, 46].

Study of change in the magnetic interaction and spin relaxation mechanism upon replacing $Dy^{3+}$ ions with $Eu^{3+}$ ions in DTO spin ice compound, thus holds significant interest. Both the compounds DTO and ETO show single ion spin freezing around 16 K and 35 K respectively. In the present investigation, we have carried out extensive analysis of $Dy_{2-x}Eu_xTi_2O_7$ (DETO) compounds (0 ≤ x ≤ 2.0) by XRD, Raman studies, DC and AC susceptibility measurements. Our data indicate that $Eu^{3+}$ ion has been substituted $Dy^{3+}$ ion systematically and it also enhances antiferromagnetic interaction. With further doping of Eu ions on Dy sites results the formation of wasp-waisted type hysteresis loop in magnetization measurement. The spin ice state exists almost throughout the whole series of DETO samples. It has also been found that spin freezing ($T_f$) is disappeared for (0 < x ≤ 1.0) samples but for (x ≥ 1.5) samples, spin freezing ($T_s$) reappeared at higher temperature in zero magnetic field. This strange behaviour may be attributed to Dy-Eu spin interactions and also to change in the crystal field due to the presence of Eu in Dy sites.

## 2. Experimental detail

$Dy_{2-x}Eu_xTi_2O_7$ (0 ≤ x ≤ 2.0) polycrystalline samples were prepared by conventional solid state reaction method. High purity (99.999%) starting materials for the synthesis were $Dy_2O_3$, $Eu_2O_3$ and $TiO_2$. Same synthesis process has been acquired to synthesize the material as reported earlier [47]. All DETO samples were confirmed to be a single phase pyrochlore without any chemical impurity phase from powder X-ray diffraction (XRD) at room temperature. XRD of samples were characterized by Rigaku Miniflex II X-ray diffractometer using Cu Kα radiation. The DC and AC magnetic measurement of each sample was performed using a Quantum Design magnetic property measurement system (MPMS) super conducting quantum interference device (SQUID) magnetometer. Raman spectroscopic measurements were carried out at room temperature, using Renishaw Micro Raman Spectrometer with a solid state laser of wavelength 532 nm.

## 3. Results and Discussions

### 3.1 Stability and Structural Analysis

The ionic radius ratio of $A^{3+}$ and $Ti^{4+}$ cations is the key factor to determine the stability of $A_2Ti_2O_7$ pyrochlore. For pyrochlore structure, value of $r(A^{3+})/r(Ti^{4+})$ (R) lies between 1.46 -1.78. For R< 1.46, the structure becomes a defect fluorite and a perovskite layered structure forms for R>1.78 [48]. We have calculated the stability determining factor of DETO samples by the following equation [49]

$$\frac{r(A^{3+})}{r(Ti^{4+})} = \frac{(1-x)r(Dy^{3+}) + xr(Eu^{3+})}{r(Ti^{4+})} \qquad (1)$$

Using eq. (1), we have determined the ionic radius ratio for each of sample which is arranged in Table 1. Here we have used the Shannon ionic radius of $Dy^{3+}$, $Eu^{3+}$, $Ti^{4+}$ ions which are 1.027 Å, 1.066 Å, 0.605 Å respectively. The value of ionic radius ratio for all samples is in between 1.46-1.78, which is a clear indication of pyrochlore structure acquired by all the samples. As the Eu content enriches in DETO samples, the $r(A^{3+})/r(Ti^{4+})$ ratio increases linearly which in turn results in enhancement of disorderness in the structure.

The room temperature X-ray diffraction data of each specimen of DETO sample confirms single phase face centered cubic pyrochlore structure with space group Fd-3m which has been displayed in inset (i) of figure 1. The phase purity, homogeneity and unit cell dimension of all powder samples were verified by Rietveld analysis of XRD data using FULLPROF program [50]. Inset (ii) of figure 1 shows the Rietveld refinement pattern of (x=1.0) sample. Evaluated lattice parameters of (x=0 and x=2.0) samples are in good agreement with earlier reports [47, 51]. The variation of lattice constant with Eu concentration is shown in figure 1. The graph shows that the value of lattice constant increase linearly with Eu concentration obeying the Vegard's law except slight deviation at x=1.9 (figure 1). This deviation from Vegard's law for x = 1.9 might be due to Dy (Eu)-Ti antisite disorder.

3.2 Raman Spectroscopy Analysis

Raman spectroscopy is very useful for structural study because it provides surface sensitive structural information of solid compounds. Raman study gives the information of interactions between spin-phonon, phonon-phonon and electron-phonon. It also tells the disorderness and phonon-anharmonicity in samples. Figure 2 shows the Raman spectra of DETO samples which recorded in the range 100 – 1000 $cm^{-1}$. For a geometrically frustrated pyrochlore $A_2B_2O_6O^{'}$ structure, there are total six Raman active modes ($A_{1g}+E_g+4F_{2g}$) and seven infrared active modes ($7F_{1u}$) as suggested by factor group theory [52]. Based on earlier reports, we have labeled all six Raman modes in figure 2 [53-55]. The most intense peak at ~ 305 $cm^{-1}$ contains two Raman modes ($F_{2g}+ E_g$) in which $F_{2g}$ mode attributes to O-Dy(Eu)-O bending modes and $E_g$ mode corresponds to O- sublattice modes. The second intense peak at ~ 518 $cm^{-1}$ is named to $A_{1g}$ mode which is related to the stretching of Dy (Eu)-O stretching bonds. The lowest frequency band near 215 $cm^{-1}$ is assigned to $F_{2g}$ mode. This first assigned $F_{2g}$ mode is associated to O'sublattice vibration as shown in figure 2. The rest two $F_{2g}$ Raman modes are located at ~ 445 cm$^{-1}$ and ~ 581 cm$^{-1}$ respectively [56].

Inset of figure 2 shows the variation of four Raman modes with x value in $Dy_{2-x}Eu_xTi_2O_7$. From the figure, it is clear that all the four modes ($F_{2g}$ ~ 215cm$^{-1}$, $F_{2g+Eg}$ ~ 305cm$^{-1}$, $F_{2g}$ ~ 445cm$^{-1}$, $A_{1g}$ ~ 518cm$^{-1}$) show red shift i.e. softening of the modes with increasing Eu content. The location of Raman active modes depends on the band strength (bond length) and ionic mass. In DETO samples, lattice constant values are increasing with increasing Eu concentration. Hence, the phonons should shift to lower frequencies. In our system, all four phonon modes shift towards lower frequency as shown in inset of figure 2. Although, a violation has been found in lattice constant variation for x = 1.9 sample from XRD data. But, the systematic shift in all phonon modes with Eu concentration (x in $Dy_{2-x}Eu_xTi_2O_7$) confirms that $Dy^{3+}$ ions are substituted by $Eu^{3+}$ ions.

## 3.3 Magnetization Analysis

The inverse DC magnetic susceptibility ($1/\chi$) vs. temperature ($T$) plot of all DETO samples is shown in figure 3a. Linear curves are found for x = 0, 0.50, 1.0, 1.5 samples as they follow CW law in whole temperature range whereas a sharp down turn is found for x = 1.8, 1.9, 2.0 samples at lower temperature. As, the low temperature magnetization data do not follow the standard CW law[ $\chi = C/(T - \theta_{CW})$, where $C$ is Curie-constant and $\theta_{CW}$ is Curie-Weiss temperature] for x = 1.8, 1.9 samples. For ETO compound, CW law also did not fit due to presence of strong plateau region (20-90 K) as shown in inset of figure 3b. This plateau region is attributed to crystal-field effect [43]. This unusual nature of $\chi^{-1}(T)$ for x = 1.8, 1.9, 2.0 samples indicates the presence of other interactions viz. exchange interactions, dipolar exchange interactions etc, along with crystal-field interactions [57]. To estimate the contribution of these magnetic interactions (viz. exchange interactions, dipolar exchange interactions), we have fitted the susceptibility data with the following equation [43];

$$\chi = C\left[\left(\frac{1}{T}\right) + \left(\frac{\theta_{CW}}{T^2}\right)\right] \qquad (2)$$

The graph between $\chi T$ vs. $1/T$ is displayed in figure 3b. We have extracted the values of $\theta_{CW}$, effective magnetic moment ($\mu_{eff}$), exchange interaction energy ($J_{nn}$) and dipolar interaction energy ($D_{nn}$) from linear fit on $\chi T$ vs. $1/T$ data for (2-5 K) temperature range. $D_{nn}$ values are obtained from $D_{nn} = \frac{\mu_0 \mu_{eff}^2}{4\pi r_{nn}^3}$ where, 'a' is the lattice constant of the unit cell of the system and $r_{nn}$ is the distance between a $A^{3+}$ ion at (0, 0, 0) and its nearest neighbour at (a/4, a/4, 0). Only classical interactions are considered here, therefore, classical exchange interaction energy ($J^{cl}$) was determined by the formula, $J^{cl} = S(S+1)J_{nn}$ and $J_{nn} = 3\theta_{CW}/zS(S+1)$ [here $z = 6$ is the co-ordination number][43]. All the obtained values are shown in Table 2. The enhancement in exchange interaction energy and $\theta_{CW}$ values, decrease in $D_{nn}$ values for x = 1.8, 1.9, 2.0 samples confirms the increase of antiferromagnetic interactions.

We have also extracted the $\mu_{eff}$ values of $Dy_{2-x}Eu_xTi_2O_7$ samples from inverse CW law fit at high temperature range (100-300 K). As the magnetic moment of both Dy and Eu ions are responsible for total magnetization value, the high temperature theoretical paramagnetic moment of all the samples can be calculated by equation, $\mu^2_{eff} = x\mu^2_{Dy} + y\mu^2_{Eu}$. Here, $x$ and $y$ are number of Dy and Eu atoms per f.u. respectively and $\mu_{Dy}$ and $\mu_{Eu}$ are the effective magnetic moment of Dy and Eu ions respectively. The experimental effective magnetic moment which was derived from CW fit by using formula, $C = \frac{N\mu^2_{eff}}{3k}$ where, $N$ is Avogadro's number and $k$ is Boltzmann constant. Variation of effective magnetic moments (both theoretical and experimental) with x value is shown in inset of figure 3c. The experimental magnetic moment value is comparable to theoretical value for x = 0, 0.50, 1.0, 1.5 samples. However, for x = 1.8, 1.9 samples, the theoretical value of $\mu_{eff}$ /Dy (Eu) is larger than the experimental one. This result suggested the strong magnetic correlations between Dy and Eu spins.

The values obtained from CW fit are very sensitive to selected temperature range [58]. In the present case, CW fit also described the data well in 2-20 K temperature range. This low temperature range is appropriate for the lowest crystal-field excitations. From the inverse linear CW fit at low temperature, we have derived the $\theta_{CW}$ values for all samples which are shown in figure 3c. These $\theta_{CW}$ are significantly positive for x = 0, 0.50, 1.0, 1.5 samples, whereas for x = 1.8, 1.9, 2.0 samples, $\theta_{CW}$ values becomes negative.

The magnetic field dependent magnetization (*M-H*) curves of ($Dy_{2-x}Eu_xTi_2O_7$) samples at 2 K are shown in figure 4a. All the curves become saturated with nearly same saturation magnetization value [Ms = 5 $\mu_B$/Dy]. The parent DTO compound is also saturated at ~ 5 $\mu_B$/Dy, which is almost half of its theoretical value. Single ion anisotropy and powder averaging effect cause to saturate the magnetic moment of DTO at ~ 5 $\mu_B$/Dy. Similar behaviour is observed for all doped samples. Saturation in magnetization around half of its theoretical value suggested that Dy spins induce the Ising anisotropy. As shown in figure 4b, interesting *M-H* behaviour is observed for x = 1.8, 1.9 samples. For these samples, the coercivity becomes minimum i.e., magnetization value at very low magnetic field becomes small. However, a loop is observed at high magnetic field. Such type of hysteresis loop is typically known as wasp-waisted hysteresis (WWH) loop [59, 60]. Such M-H loops are commonly observed in mixed magnetic states that might develop due to change in spin exchange coupling between the cations [61, 62]. It is also supposed that WWH loop arises from presence of different domain states, coercive fields and chemical disorder [63-66]. Basically, chemical disorder in a system results in frustration in magnetic states which is favored by competitions between antiferromagnetic and ferromagnetic interactions.

Moreover, in formation of WWH loop major roles are played by the dipolar interaction ($D_{nn}$) and magneto-crystalline anisotropy ($K_1$) [67]. As has been mentioned above, the $D_{nn}$ has been determined by fitting M(T) data with the eq (1). It is observed that $D_{nn}$ is maximum for x=1.8 which is consistent with WWH loop value. It is worthwhile to mention that WWH could be found as an effect of the coexistence of multi-domains and single-domains. But the present samples are bulk and therefore possibility of the existence of single domain can be discarded. Furthermore, if this was the origin then squareness should have been there in the M(H) loop. But in the present case, such M(H) loop was not observed. Similar behavior has been observed in $CoFe_2O_4$ nano powder [67]. Further, to find out the various components causing WWH loop for x = 1.8, 1.9 compounds, we have plotted the difference of magnetization ($\delta M$) between magnetization values with increasing and decreasing magnetic field, particularly, for first quadrant of M-H curves (shown in inset of Figure 4c). For both samples, $\delta M$ curves show two anomalies (change in slope) near around 0.01 T and 0.6 T magnetic field. We have plotted the double derivative of $\delta M$ curves ($d^2\delta M/dH^2$) with H for x = 1.8, 1.9 samples to reveal the role of different coercivities as shown in figure 4c. Both the $d^2\delta M/dH^2$ curves instead of showing usual single peak exhibit two peaks centered on around 0.0055 T and 0.55 T respectively. The existence of two peaks indicates the presence of two switching field distributions, which is resulting from the dipolar field interaction and easy plane anisotropy [67]. Similar analysis has been performed to know the origin of WWH loop for 1D $Nd_{0.1}Bi_{0.9}FeO_3$ (NBFO) nano-tubes [68] and for $CoFe_2O_4$ nano-powder [67]. In NBFO, two peaks (around 0.03 T and 1.7 T) have been observed in $\delta M/dH$ curve. Low field peak is associated to low coercivity (soft component), where as high field peak is attributed to hard component [68].

Furthermore, we have also calculated the value of anisotropy constant ($K_1$) for x = 1.5, 1.8, 1.9 samples. We have used the law of approach to saturation model [69] and fitted the data with the equation;

$$M = M_s \left[1 - \frac{8}{105} \frac{K_1^2}{\mu_o^2 M_s^2} \frac{1}{H^2}\right] + kH, \qquad (3)$$

where $M$ is magnetization, $M_s$ is saturation magnetization, $K_1$ is first order cubic anisotropy coefficient, $H$ is applied magnetic field, and $k$ is the forced magnetization coefficient and constant 8/105 deals to first order cubic anisotropy of random polycrystalline samples [70]. The descending magnetization data of first quadrant of M-H curves has been fitted by eq. 3 above 1 T applied magnetic field at 2K as shown in inset of figure 4a. We have extracted the values of $K_1$, $M_s$ and $k$ for all three samples from the fit. The obtained $M_s$ values of x = 1.5, 1.8, 1.9 samples from the fitting of experimental data by eq. 3 are in agreement with magnetization value measured at applied 2 T magnetic field. The extracted $K_1$ values are listed in Table 3.

In order to examine the spin freezing and to investigate time dependent magnetic properties of doped $Dy_{2-x}Eu_xTi_2O_7$ compounds, AC susceptibility measurements have been performed. Figure 5a shows the real part of AC susceptibility of x = 0.0, 0.5, 1.0 samples and inset (i) of figure 5a shows the $\chi'(T)$ of x = 1.5, 1.8, 1.9 systems. Whereas, figure 5b shows the corresponding imaginary part of AC susceptibility at 500 Hz frequency and zero magnetic field. For pure DTO compound (x = 0.0), two dips are appeared in $\chi'(T)$ curve around 3 K (indicated by $T_i$) and 16 K (indicated by $T_f$) corresponding two peaks in $\chi''(T)$ are also visible. The drop around 3 K ($T_i$) is well known spin ice freezing transition [34, 35] and the second drop around 16 K ($T_f$) in $\chi'(T)$ curve is associated to single ion freezing due to $Dy^{3+}$ ion [36]. The low temperature $T_i$ spin ice freezing is present in Eu doped x = 0.5, 1.0, 1.5 samples. This indicates the robustness of spin ice state in DETO, inspite of the increase in antiferromagnetic interactions between spins. The similar observations were also found in stuffed spin ice $Ho_2(Ti_{2-x}Ho_x)O_{7-x/2}$ [71] and $Dy_2(Ti_{2-x}Dy_x)O_{7-x/2}$ [72]. From inset (i) of figure 5a and 5b, it has been shown that $T_i$ peak is not observed for x = 1.8, 1.9 samples. However, observed finite value of AC susceptibility for x = 1.8, 1.9 samples is indicating that $Dy^{3+}$ spins are still fluctuating at lower temperature (below 2 K). These low temperature outcomes are expected for quantum spin ice [73]. Moreover, $Dy^{3+}$ spins are Ising type in all samples. Thus, they have ice-like configuration. These all results suggest that DETO samples remain into a spin ice state. The $\chi'(T)$ and $\chi''(T)$ curves exhibit paramagnetic behaviour at high temperature with H = 0 T magnetic field for x = 0.5, 1.0 compounds i.e. $T_f$ spin freezing is disappeared. Unusual nature has been found for x = 1.5, 1.8, 1.9 samples, in which high temperature spin freezing has reemerged at T > 18 K [inset (i) of figure 5a and 5b]. Similar re-entrance of high temperature spin freezing was noticed for diluted spin ice $Dy_{2-x}M_xTi_2O_7$ (M = Y, Lu), in which $T_f$ peak is subdued for low levels of dilution of non – magnetic $Y^{3+}$ and $Lu^{3+}$ ions. However, this $T_f$ peak is re-emerged for high dilution (x > 0.4) towards higher temperature. This re-entrance spin freezing is dominated by single ion effect and attributed to change in crystal field levels along with quantum mechanical and thermal processes [36]. However, the scenario is different for our $Dy_{2-x}Eu_xTi_2O_7$ series. Because, $Eu_2Ti_2O_7$ (x = 2.0) also shows a single ion spin freezing transition ($T_s$) at 35 K which is shown in inset (ii) of figure 5a and 5b [42]. This re-entrant high temperature spin freezing for x = 1.5, 1.8, 1.9 might be associated to $Eu^{3+}$ spins.

To find out the origin of re-entrant spin freezing for Eu rich samples, DC magnetic field of 2 T was applied. An applied magnetic field let up the dynamical process and due to this there is an enhancement in χ' and χ" signals. Figure 6 shows the AC susceptibility [upper panel $\chi'(T)$ and lower panel $\chi''(T)$] of all doped samples at H = 2 T. For x = 0.5, 1.0 samples, a clear transition ($T_f$ marked by brown arrow) is appeared in $\chi'(T)$ part above 16 K, corresponding a peak in $\chi''(T)$ [figure 6a, 6b]. This field dependent data is in contrast to that of H = 0 T as no high temperature peak observed is there. Furthermore, for x = 1.5 compound, an additional drop ($T_s$ indicated by green arrow) is observed in $\chi'(T)$ around 23.5 K and corresponding peak is appeared in $\chi''(T)$ along with $T_f$ [figure 6c]. Similar new additional peak ($T_s$) has also been observed for x = 1.8, 1.9 at higher temperature (T > 23.5 K) [in figure 6d, 6e]. As a matter of fact, two peaks ($T_f$ and $T_s$) are observed for (x > 1.0) samples on applying H = 2 T magnetic field. The emergence of $T_s$ peak in $Dy_{2-x}Eu_xTi_2O_7$ is similar to that reported in $Dy_xTb_{2-x}Ti_2O_7$, $Dy_{2-x}Gd_xTi_2O_7$ and $Dy_{2-x}Yb_xTi_2O_7$ [44, 74, 75].

Both $T_f$ and $T_s$ transitions are thermally activated. Therefore, to confirm the nature of transition, we have fitted the frequency dependence of $T_f$ and $T_s$ peaks for all doped samples by Arrhenius Law, $f = f_0\ exp(E_a/k_bT)$ where $E_a$ is the activation energy for spin fluctuation, $f_0$ is a measure of the microscopic limiting frequency in the system and $k_B$ is the Boltzmann constant. Figure 7a shows the Arrhenius fit. The $T_f$ peak in all doped samples correspond to single ion freezing for $Dy^{3+}$ ions. As the extracted thermal energy values of doped DETO samples which are linked to crystal field splitting levels are qualitatively similar to DTO, $T_f$ peak is identified as the single-ion freezing of $Dy^{3+}$ spins. The calculated values of $f_0$ for $T_f$ peak are varied from 0.05 to 1.06 GHz. The extracted values of $E_a$ for $T_f$ and $T_s$ peaks are shown in figure 7b. The variation of $T_f$ and $T_s$ peaks with x value in $Dy_{2-x}Eu_xTi_2O_7$ is shown in figure 7c. The $T_s$ peak has emerged only for those samples which contain large quantity of $Eu^{3+}$ ions ($1.5 \leq x \leq 2.0$) or only $Eu^{3+}$ ions. Derived $E_a$ value from Arrhenius fit of $T_s$ peak (originated from $Eu^{3+}$) is increased from 277.8 K (x = 1.5) to 327.08 K (x = 1.9), whereas for pure ETO, $E_a$ was reported to 339 K [42] and $f_0$ value ranges from 0.05 to 0.5 GHz. The obtained $E_a$ value from the fitting of $T_s$ peak is close to value assigned for single ion freezing due to $Eu^{3+}$ spins which confirms that $T_s$ peak is originated because of $Eu^{3+}$ ions. The variation in crystal field levels is due to Dy- Eu interactions, causes the systematic alteration in $E_a$ value of $T_f$ peak for ($1.5 \leq x < 2.0$) samples. It can be further verified from the fact that crystal field effect is closely related to structural parameters like lattice constant and atomic position of oxygen atom surrounding the rare earth (A) ion [76]. From structure study, it has been found that on increasing the concentration of $Eu^{3+}$ ions, lattice constant of DETO system was increased. This indicates that doping of $Eu^{3+}$ content alters the crystal field levels. However, it is expected that activation energy value ($E_a$) should decrease on increasing the lattice constant. But, in contrast enhancement in $E_a$ value was observed for present case. This nature specifies the influence of antisite disorder which causes the change in crystal-field, consequently change in energy barriers. Withal, $T_s$ moves to higher temperature side with increasing Eu dopant which supports the increase value of $E_a$ as observed in figure 7c [77].

From figure 7c, it has been observed that position of $T_f$ peak (associated to $Dy^{3+}$ spins) is deviated from x = 1.5 sample, where $T_s$ peak (associated to $Eu^{3+}$ spins) in introduced in DETO samples. In DETO pyrochlore (corner sharing tetrahedron structure), it is assumed that $Dy^{3+}$ spins and $Eu^{3+}$ spins randomly occupy the lattice of tetrahedron. For (x ≥ 1.5) samples, only tetrahedron of $Eu^{3+}$ spins appears, whereas for (x <1.5), only tetrahedron of $Dy^{3+}$ spins emerges. However, $Dy^{3+}(Eu^{3+})$ lies in the neighborhood coordinated with six $Eu^{3+}(Dy^{3+})$ spins. Consequently, $T_f$ peak is deviated from x = 1.5. This argument further suggested that $T_s$ transition is associated to spin freezing of $Eu^{3+}$ ions.

However, from the above discussion it is clear that x =1.8 and 1.9 samples exhibit WWH loop and also spin frustration. The spin frustration is also observed in x=1.5 samples (figure 6c) but it does not show wasp-waisted hysteresis loop. Therefore, spin frustration is not the origin in the present case as has been discussed in the $Fe_2MnGa$ system [66]. Furthermore, as it has already been discussed that both x=1.8 and x=1.9 samples have significant dipolar field and anisotropy exchange interaction whereas in x=1.5 sample existence of no dipolar field is found. Therefore, existence of dipolar field and anisotropy exchange interaction is the origin of WWH loop in the present case.

Our observations including the compositions, frequency and magnetic field dependence of $T_f$ and $T_s$ peaks indicate that doping of $Eu^{3+}$ ions at A site in DETO samples result the alteration in crystal-field levels. The Eu-Dy spins interaction changes this crystal field effect. Unlike other hybrid pyrochlore compounds, $Dy_{2-x}Eu_xTi_2O_7$ may be considered as combination of DTO (spin ice) and ETO. However, further experimental and theoretical analysis is needed for a clear vision of relaxation mechanism which may be helpful for understanding the unrevealed physics of the system.

## 4. Conclusion

The Raman effect and magnetization study have been made on $Dy_{2-x}Eu_xTi_2O_7$ pyrochlore. From the Raman effect study, it is clear that Dy is replaced by Eu. M(T) data of low Eu content samples can be fitted with the Curie-Weiss law and rich Eu containing samples can be fitted with the high temperature series expansion equation. Large Eu induces the dipolar field and anisotropy exchange interaction. The coexistence of these dipolar field and anisotropy exchange interaction induce wasp-waisted hysteresis loop in $Dy_{2-x}Eu_xTi_2O_7$ system. AC susceptibility measurement indicates the existence of both single ion spin freezing (due to $Dy^{3+}$ and $Eu^{3+}$ ions) in Eu rich samples. It is also observed that crystal field effect is closely related to structural parameters like lattice constant and atomic position of oxygen atom surrounding the rare earth (A) ion.

## Acknowledgement

The authors are thankful to the CIF, Indian Institute of Technology (BHU) for providing the facility of magnetic measurements. PS and VKG are also expressing their gratitude to UGC India for providing fellowship. The financial support by DST-FIST to the department is also gratefully acknowledged.

**Table Caption**

**Table (1):** The ionic radius ratio of DETO series compounds.

**Table (2):** Extracted Curie–Weiss temperature, classical exchange energy, dipolar interaction energy and calculated magnetic moment of (x = 1.8, 1.9, 2.0) samples by high temperature series expansion fit [2-5 K].

**Table (3):** Extracted anisotropy constant ($K_1$) values of x = 1.5, 1.8, 1.9 samples.

**Figure Captions:**

**Figure (1)**: The lattice constant (Cubic Structure) of $Dy_{2-x}Eu_xTi_2O_7$ as a function of x value. Inset **(i):** X-ray powder diffraction pattern for the $Dy_{2-x}Eu_xTi_2O_7$ samples. Inset **(ii):** Rietveld refinement for the $Dy_{1.0}Eu_{1.0}Ti_2O_7$ sample.

**Figure 2(a):** Raman spectra of the $Dy_{2-x}Eu_xTi_2O_7$ samples at 300 K. Inset: Variation of four active phonon modes as a function of x in $Dy_{2-x}Eu_xTi_2O_7$ samplesalong with the straight horizontal dashed lines for reference.

**Figure 3(a):** Inverse DC susceptibility vs. T curve of $Dy_{2-x}Eu_xTi_2O_7$ compounds with inverse CW fit at (100-300 K). **(b):** High temperature series expansion fit for $Dy_{2-x}Eu_xTi_2O_7$ (x = 1.8, 1.9, 2.0) samples. Inset: $\chi$ vs. T curve of x = 2.0 sample with standard CW fit. **(c):** Variation of derived CW temperature with x value of $Dy_{2-x}Eu_xTi_2O_7$ series from (2-20 K) CW fit. Inset: Variation of calculated effective magnetic moments with Eu content (x) derived from inverse CW fit at high temperature (100-300 K) for DETO samples.

**Figure 4(a)**: M-H curves at 2 K for all DETO samples. Inset: Descending first quadrant M-H curve for x = 1.5, 1.8, 1.9 samples with LA fit above 1 T **(b):** Zoom part of M-H curve for x = 1.8, 1.9 samples at 2 K. **(c):** Variation of $d^2\delta M/dH^2$ curve with applied H at 2 K for x = 1.8, 1.9 samples. Inset: $\delta M$ between ascending and descending portions of M-H curves for x = 1.8, 1.9 samples..

**Figure 5(a):** $\chi'(T)$ of x = 0.0, 0.5, 1.0 for $f$ = 500 Hz at zero applied DC field. Inset **(i):** $\chi'(T)$ of x = 1.5, 1.8, 1.9 for $f$ = 500 Hz at zero applied DC field. Inset (ii): $\chi'(T)$ of x = 2.0 for $f$ = 500 Hz at zero applied DC field. **(b):** $\chi''(T)$ of x = 0.0, 0.5, 1.0 for $f$ = 500 Hz at zero applied DC field. Inset **(i):** $\chi''(T)$ of x = 1.5, 1.8, 1.9 for $f$ = 500 Hz at zero applied DC field. Inset **(ii):** $\chi''(T)$ of x = 2.0 for $f$ = 500 Hz at zero applied DC field.

**Figure (6):** $\chi'(T)$ (upper panel) and $\chi''(T)$ (lower panel) of $Dy_{2-x}Eu_xTi_2O_7$ compounds at applied field of 2 T. **(a):** $Dy_{1.5}Eu_{0.5}Ti_2O_7$, **(b):** $Dy_{1.0}Eu_{1.0}Ti_2O_7$, **(c):** $Dy_{0.5}Eu_{1.5}Ti_2O_7$, **(d):** $Dy_{0.2}Eu_{1.8}Ti_2O_7$, **(e):** $Dy_{0.1}Eu_{1.9}Ti_2O_7$. Marked by arrow are: Single ion spin freezing peak of $Dy^{3+}$ ions ($T_f$) and Single ion spin freezing peak of $Eu^{3+}$ ions ($T_s$).

**Figure 7(a):** The Arrhenius Fit of ($T_f$) peak for (x = 0.5, 1.0, 1.5, 1.8, 1.9) compounds and Arrhenius Fit of ($T_s$) peak for (x = 1.5, 1.8, 1.9) compounds.**(b):** Variation of derived $E_a$ values of $T_f$ and $T_s$ peaks with x value of $Dy_{2-x}Eu_xTi_2O_7$.**(c):** Variation of $T_f$ and $T_s$ peaks with x value of $Dy_{2-x}Eu_xTi_2O_7$.

**Tables**

**Table (1)**

| Sample | x = 0.0 | x = 0.5 | x = 1.0 | x = 1.5 | x = 1.8 | x= 1.9 | x = 2.0 |
|---|---|---|---|---|---|---|---|
| $r(R^{3+})/r(Ti^{4+})$ | 1.69 | 1.714 | 1.73 | 1.75 | 1.755 | 1.758 | 1.762 |

**Table (2)**

| Sample | $\theta_{CW}$ (K) | $J^{cl}$ (K) | $D_{nn}$ (K) | $\mu_{eff}$ ($\mu_B$) |
|---|---|---|---|---|
| $Dy_{0.2}Eu_{1.8}Ti_2O_7$ | -0.0826 | -0.0413 | 0.126 | 3.088 |
| $Dy_{0.1}Eu_{1.9}Ti_2O_7$ | -0.099 | -0.0496 | 0.06037 | 2.13 |
| $Eu_2Ti_2O_7$ | -1.35 | -0.6742 | 0.00609 | 0.679 |

**Table 3**

| S.N. | Sample | Anisotropy Constant ($K_1$) (J/m$^3$) |
|---|---|---|
| 1. | $Dy_{0.5}Eu_{1.5}Ti_2O_7$ | $1.132 \times 10^5$ |
| 2. | $Dy_{0.2}Eu_{1.8}Ti_2O_7$ | $4.603 \times 10^4$ |
| 3. | $Dy_{0.1}Eu_{1.9}Ti_2O_7$ | $2.649 \times 10^4$ |

**Figures:**

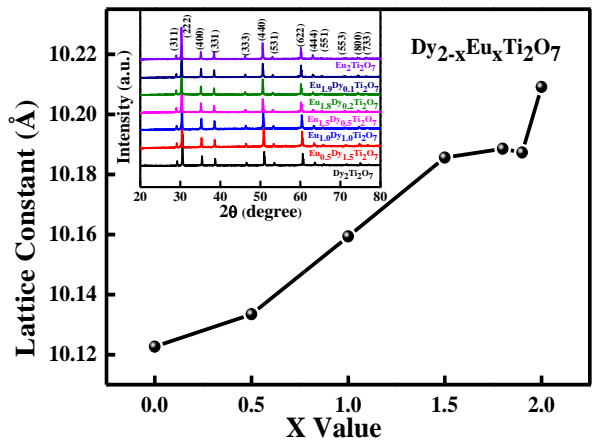

**Figure 1**

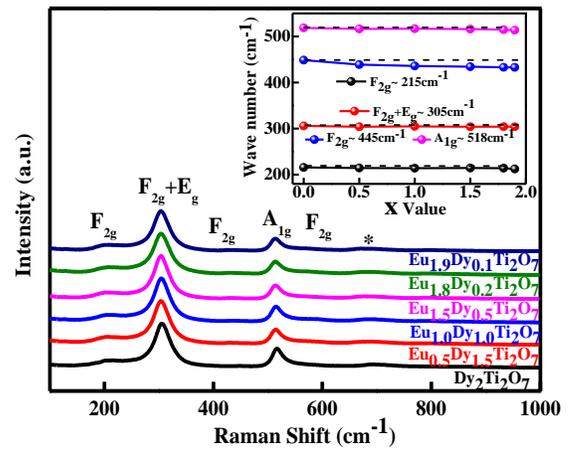

**Figure 2**

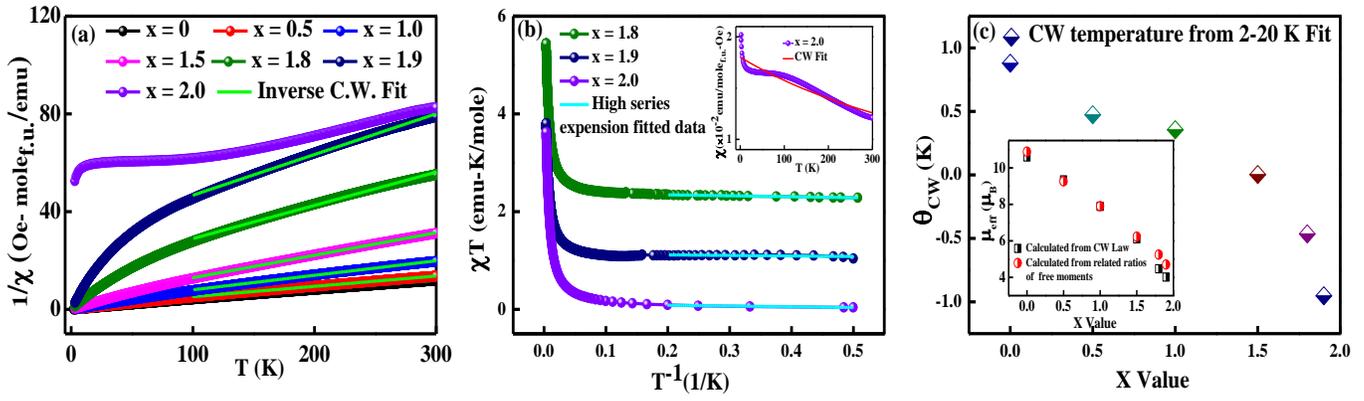

**Figure 3**

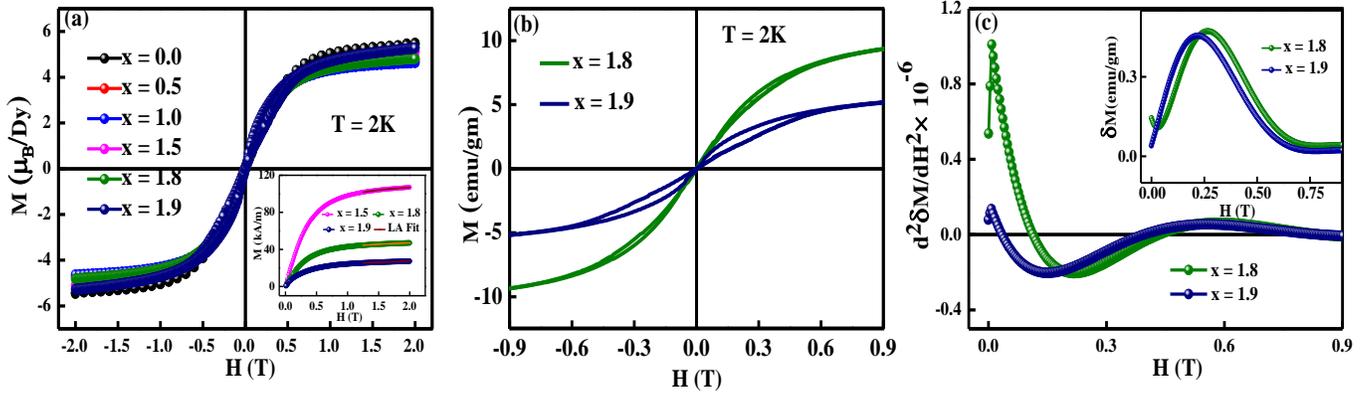

**Figure 4**

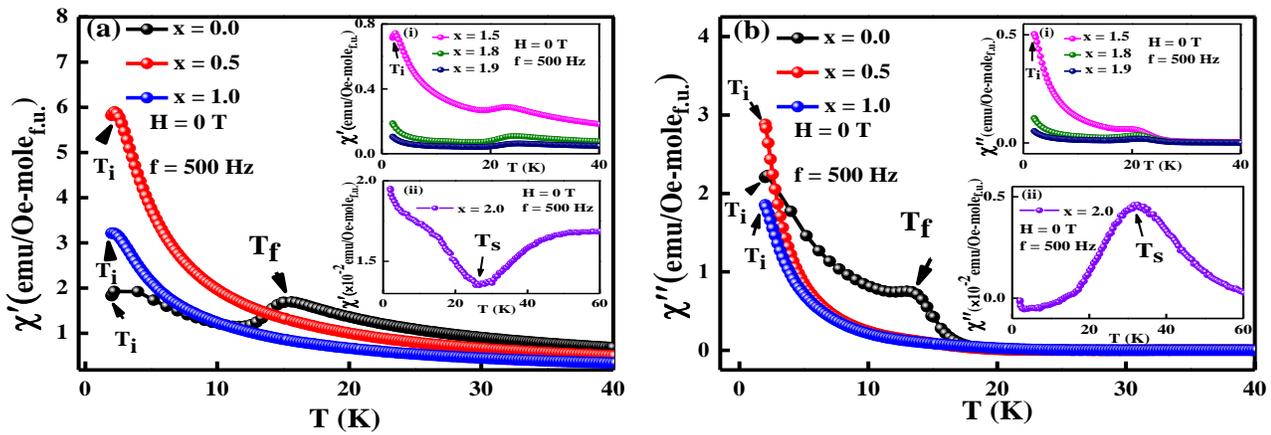

**Figure 5**

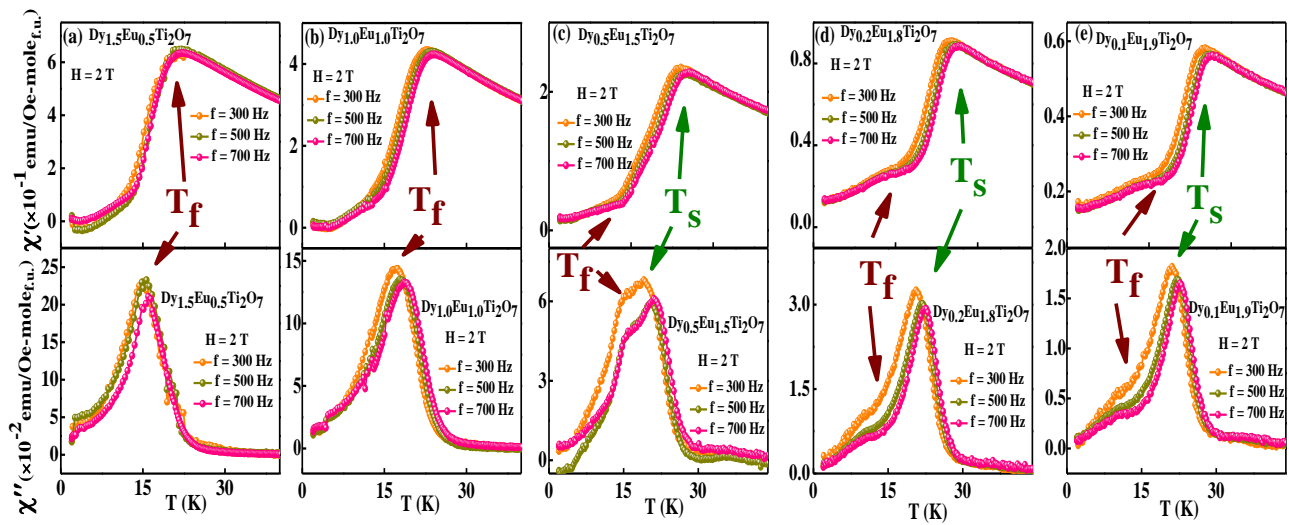

**Figure 6**

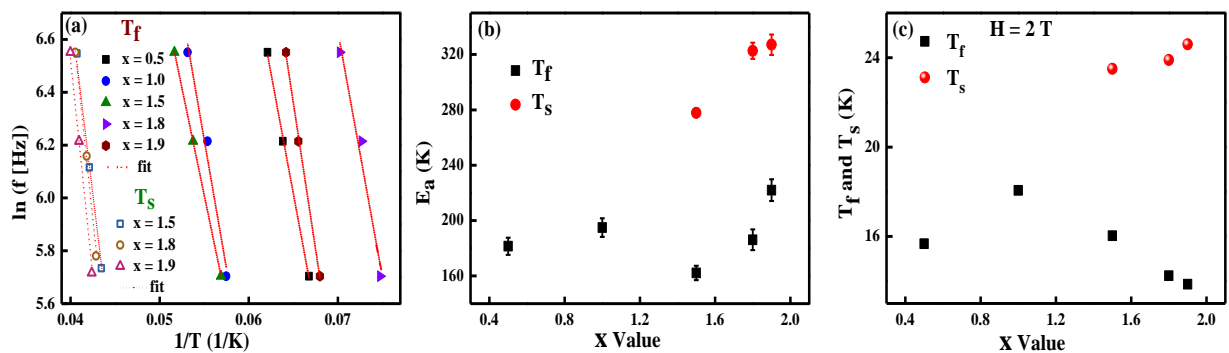

**Figure 7**